# NEW NARROW N*(1685) RESONANCE: REVIEW OF OBSERVATIONS

Viacheslav Kuznetsov[1a]

[1] Petersburg Nuclear Physics Institute, 188300, Gatchina, Russia

**Abstract.** The recent Review of Particle Physics [1] includes a new narrow N*(1685) resonance. Its properties, the narrow width ($\Gamma < 25$ MeV) and the strong photoexcitation on the neutron, are unusual. The paper reviews experimental data which have led to the evidence for this resonance. Alternative explanations of the phenomenon at W ~ 1.685 GeV, namely the interference of well-known resonances and the sub-threshold meson-baryon production, are discussed as well.

Over decades the problem of ``missing" resonances remains one of the major challenges in the domain of hadronic physics. Many baryon resonances predicted on the basis of Constituent Quark Model (CQM), were not found in experiment. Although extensive programs to search for these resonances are underway at several facilities (CBELSA/TAPS, JLAB, A2@MaMiC *etc.*), it seems timely to assume that the ``missing" resonances may not exist while the revision of theoretical predictions is needed.

In the mean-field approach (MPA) [2-4] baryons are treated as multiquark systems stored in mean field similar to large-A nuclei made of protons and neutrons. Remarkably, this approach well describes the spectrum of known resonances. ``…There are no resonances (below 2 GeV) from the PDG table which are left unaccounted for, and … no extra states from the theory side except (one) $\Delta^+_{3/2}$..." (citation from [3]). As by-product, MFA predicts the existence of long-lived exotic baryons.

Therefore, search for exotic states is quite important to establish the validity of two models (CQM and MFA) and the connection between them. In this context the evidence for a new narrow N*(1685) resonance is of potential importance. Historically, the first observation of the unusual peculiarity at W~1.68 Gev stems from the study of $\gamma n \rightarrow \eta n$ and $\gamma p \rightarrow \eta p$ reactions at GRAAL. In 2002 the GRAAL Collaboration reported the sharp rise in the ratio of the η photoproduction cross sections on the neutron and the proton at the photon energy ~1 GeV [5]. Further a bump-like structure on the neutron at $E_\gamma$~ 1.03 GeV was found [6,7]. This structure was not (or poorly) seen on the proton. Somewhat later CBELSA/TAPS [8,9] and LNS-Sendai [10] collaborations confirmed this observation. Very recently the A2@Mainz Collaboration published the precise $\gamma n \rightarrow \eta n$ data on the deuteron and $^3$He [11,12].

The peculiarity was observed as a bump in the quasi-free cross sections and as a peak in the invariant-mass spectra of the final-state η and the recoil neutron. The width of the bump in the cross sections was close to that expected for a narrow resonance smeared by Fermi motion of the target neutron. The widths of the peaks in the M(η,n) spectra were dominated by the instrumental resolutions of corresponding experiments. Later the CBTAPS/ELSA and GRAAL groups managed to reduce the effect of Fermi motion [9,13] (Fig. 1). The current width estimate is $\Gamma < 25$ MeV [9].

---

[a] Corresponding author: slava@pnpi.spb.ru

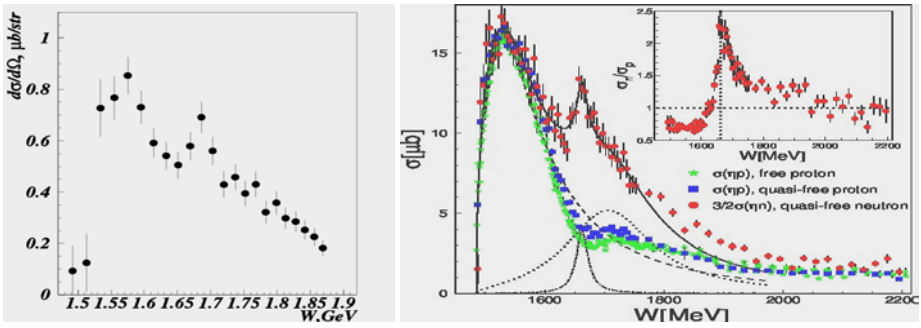

Fig.1 Recent γn → ηn cross sections from GRAAL [13] (left) and CBELSA/TAPS [9] (right).

The observation of the narrow structure in γn → ηn raised debates about its nature. One obvious explanation was the existence of a narrow resonance whose photoexcitation is suppressed on the proton and is strong on the neutron [for example, 14-16]. It was disputed by the authors of [17,18]. They suggested that the interference of well-known resonances generates this enhancement. In [19,20] the bump was explained as a possible sub-threshold meson-nucleon production (cusp-effect).

If photoexcitation of a resonance occurs on the neutron and is suppressed on the proton, such a resonance would appear in the proton cross section as a minor peak/dip structure which could be unresolved in experiment. This resonance may manifest itself in polarization observables in which its signal could be amplified due to the interference with other resonances. The revision of the of γp → ηp beam asymmetry data from GRAAL revealed an oscillating resonant structure at $W=1.685$ GeV [21,22] (Fig2, left). Further, the small dip structure was found by the A2@Mainz Collaboration in the high-precision of γp → ηp cross section data [23] (Fig.2, right). As it was shown in [24], these data may hint a narrow resonance with the mass near 1.69 GeV.

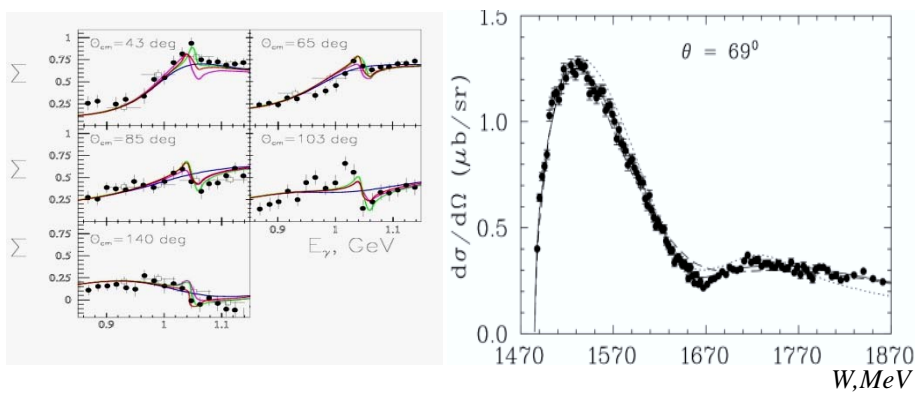

Fig.2 On the left: energy dependence of γp → ηp beam asymmetry [21,22]. On the right: one example of γp → ηp differential cross section (data from [23]).

If the narrow N*(1685) does really exist, it should manifest itself in different reactions. The peak at the same energy was found in Compton scattering on the neutron at GRAAL [25] (Fig.3, left). It was not seen (or unresolved) in the γn → γn and γn → $π^0$n [25, 26]. The small but clear peak structure at W ~ 1.686 GeV was observed in the high-precision measurement of $π^-$p → $π^-$p at the EPECUR facility [27] (Fig.3, right).

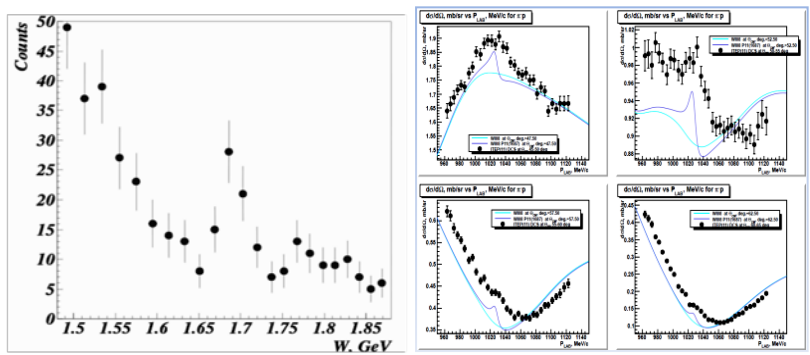

Fig.3 On the left: yield of γn → γn events [24]. On the right: π⁻p → π⁻p data from EPECUR [26].

The authors of [17,18] explained the bump in γn → ηn as the specific interference of known wide resonances ($S_{11}$(1535) and $S_{11}$(1650) [17] or $S_{11}$(1650) and $P_{11}$(1710) [18]). Here are some critical remarks:
i)   The widths of the bumps in [17, 18] seem to be wider than in the recent GRAAL and CBELSA/TAPS results (Fig.1);
ii)  It is unclear whether this interference can generate the narrow structure in γp → ηp, in particular, in the beam asymmetry (Fig.2);
iii) Unlikely this interference can generate the structure in γn → γn and π⁻p → π⁻p at the same energy since these reactions are governed by different resonances.

The idea of the cusp effect is mostly based on the model-dependent calculations by M.Doring and K.Nakayama [19]. They demonstrated that the sub-threshold virtual KΣ production may generate the bump at W~1.68 GeV in the γn → ηn cross section while KΛ generates a dip structure at the slightly lower energy. At the same time the γp → ηp cross section remains flat. The authors managed to qualitatively reproduce the enhancement in the γn → ηn cross section and the peak in the $σ_n/σ_p$ ratio. However, some points require the clarification:
i)   Can these calculations reproduce the narrow width of the peak in the recent updates from GRAAL and CBELSA/TAPS (Fig.1)?
ii)  Can the cusp effect generate the narrow structure in the γp → ηp beam asymmetry (Fig.2)?
iii) Can the cusp effect generate the bump in Compton scattering on the neutron which is an electromagnetic process?
iv)  Why this effect is suppressed in γn → π⁰n and π⁻p → π⁻p?

At present, the only explanation that accommodates all experimental findings is the existence of a narrow resonance with the following properties:
- Mass near 1.68 GeV;
- Width Γ <25 MeV;
- Isosspin 1/2;
- Strong photoexcipation on the neutron and suppressed photoexcitation on the proton;
- Suppressed decay to πN final state.

These properties are similar those expected for the second member of the exotic antidecuplet [28,29] predicted by the Chiral Soliton Model (χSM) [30]. On the other hand the decisive identification of this resonance requires more efforts, in particular, the precise determination of its width and quantum numbers. Apart from that, the predictions of MFA (the natural extension of χSM) for the antidecuplet and its members are still unclear.

The author wishes to thanks the administration of Petersburg Nuclear Physics Institute for support in attending this Conference. Discussions with M.Polyakov, B.Krushe, and H.Schmieden were quite helpful.